\documentclass[epjST]{svjour}
\usepackage{graphics}
\begin{document}
\title{A brief history of the introduction of generalized ensembles 
to Markov chain\\ Monte Carlo simulations}
\subtitle{Festschrift to the occasion of Wolfhard Janke's 60th birthday}
\author{Bernd A Berg\thanks{\email{bberg@fsu.edu}}}
\institute{Department of Physics, Florida State University, 
Tallahassee, FL 32306, USA }
\abstract{
The most efficient weights for Markov chain Monte Carlo calculations 
of physical observables are not necessarily those of the canonical 
ensemble. Generalized ensembles, which do not exist in nature but can 
be simulated on computers, lead often to a much faster convergence. In 
particular, they have been used for simulations of first order phase 
transitions and for simulations of complex systems in which conflicting 
constraints lead to a rugged free energy landscape. Starting off with 
the Metropolis algorithm and Hastings' extension, I present a minireview 
which focuses on the explosive use of generalized ensembles in the early 
1990s. Illustrations are given, which range from spin 
models to peptides.
} 
\maketitle

\section{Introduction} \label{sec_intro}

In this minireview I give a personal account of the birth of 
generalized ensemble simulations, which happened in the early 1990s.
Though essential ideas go well back into the 1970s, they were not
build on, but rediscovered around 1990. Later papers were often unaware 
of their predecessors. Nevertheless, I make here an attempt to present
developments in a chronological order. Starting with the Metropolis 
algorithm, subsequent early work on the methodology, and Hastings' 
extension to a general Markov chain Monte Carlo (MCMC) scheme, I come 
then to the introduction of generalized ensembles: Umbrella sampling 
and more than a decade later the almost simultaneous emergence of 
multicanonical (MUCA) ensembles, expanded ensembles and replica 
exchange simulations (parallel tempering).

My account of the last two methods remains rather sketchy as I was not 
involved into their development. My emphasis is on publications in which 
I participated and those which influenced my own work or crossed its 
path by one or another reason. No attempt of an ``objective'' overview 
or covering the field beyond the early developments is made. So, my 
focus is on MUCA simulations for which calculations of interface 
tensions of first order phase transitions via Binder's method were
the point of departure. Subsequently, I discuss applications to spin 
glasses and peptides (small proteins) as examples of complex systems 
with conflicting constraints. Concerning the performance, it is outlined 
that even for first order phase transitions a residual exponential 
slowing down remains. Wang-Landau sampling is recommended for 
calculating the MUCA weights. Summary and conclusion are given 
in the final section.

\section{Metropolis algorithm and Hastings' extension} \label{sec_Met}

MCMC calculations started in earnest with the 1953 paper by Metropolis,
Rosenbluth$^2$ and Teller$^2$ \cite{Met}. In their own words: ``A general
method, suitable for fast computing machines, for investigating such
properties as the equations of state for substances consisting of 
interacting molecules is described. The method consists of a modified
Monte Carlo integration over configuration space.'' It relies on 
calculating the change in energy $\triangle E$, which is caused by 
a stochastic move of a particle position: ``If $\triangle E<0$, i.e., 
if the move would bring the system to a state of lower energy, we allow 
the move and put the particle in its new position. If $\triangle E>0$, 
we allow the move with probability $\exp(-\triangle E/kT)$; i.e., we 
take a random number $\xi_3$ between 0 and 1, and if $\xi_3<\exp(-
\triangle E/kT)$, we move the particle to its new position. If $\xi_3>
\exp(-\triangle E/kT)$, we return it to its old position. Then, whether 
the move has been allowed or not, i.e., whether we are in a different 
configuration or in the original configuration, we consider that we are 
in a new configuration for the purpose of taking our averages.''

Whether to count a configuration after a rejected move again or not
caused some discussion at its time and may still be a stumbling block
for newcomers. I remember that I found the false choice in a lattice
gauge theory (LGT) Metropolis programs of a postdoc in the 1990s, who 
had been running them for quite while on supercomputers.

\subsection{Notable early developments} \label{sec_rwght}

In 1959 Salsburg et al.~\cite{Sals} made an attempt to calculate the
partition function by MCMC using a method called in the modern language
``reweighting'':
\begin{eqnarray}
  \exp(-\beta E)\ \to\ \exp(-\beta E-\triangle\beta E) = 
  \exp(-\beta'E)\,.
\end{eqnarray}
As noticed by the authors their method is restricted to very small
lattices when one wants to cover the entire $\beta$ range from 0
to $\infty$ as needed for the partition function, because for a fixed, 
possibly large, statistics and away from critical points $\triangle
\beta\sim 1/\sqrt{\rm V}$ ($V$ volume) holds for the admissible 
$\triangle \beta$ range.

In the next almost thirty years reweighting was several times forgotten
and re-discovered (for instance, McDonald and Singer \cite{McDS} used
reweighting to evaluate physical quantities over a small range of 
temperatures) until finally the time was right and the MCMC community 
had grown large enough so that reweighting never became extinct again 
after the 1988 paper by Ferrenberg and Swendsen \cite{FS}.

Continuing with the MCMC time line, in his 1963 article Glauber 
\cite{Glauber} transcends the Metropolis dynamics by introducing 
the heatbath and other dynamical approaches for the Ising model.

\subsection{Hastings} \label{sec_hast}

In Hastings own words \cite{Hastings}: ``A generalization of the 
sampling method introduced by Metropolis et al.\ (1953) is presented 
along with an exposition of the relevant theory, techniques of 
application and methods and difficulties of assessing the error in 
Monte Carlo estimates.'' Unfortunately, the mathematician Hastings 
renamed everything the chemists and physicists had introduced before: 
``Equilibrium'' to ``stationary'', ``detailed balance'' to 
``reversibility'' and ``ergodicity'' to ``irreducibility''.  

Instead of focusing on the Boltzmann distribution, Hastings constructs
a Markov chain matrix $P$ with a general stationary distribution $\pi$. 
Following his paper, the matrix $P$ is made to satisfy the detailed 
balance (reversibility) condition
\begin{eqnarray}
  \pi_i\,p_{ij} = \pi_j\,p_{ji}
\end{eqnarray}
for all pairs of states $i$ and $j$. This property ensures that 
$\sum_i\pi_i p_{ij}=\pi_j$ for all $j$, and hence that $\pi$ is 
an equilibrium (stationary) distribution of $P$. Ergodicity 
(irreducibility) of $P$ must be checked in each specific 
application. For this it is only necessary that there is
a positive probability of going from state $i$ to state $j$
in some finite number of transitions.

It is then assumed that $p_{ij}$ has the form
\begin{eqnarray}
  \pi_i\,p_{ij} = q_{ij}\,\alpha_{ij}\ (i\ne j)~~{\rm with}
  ~~ p_{ij} = 1 - \sum_{j\ne i} p_{ij}\,.
\end{eqnarray}
Here $Q=\{q_{ij}\}$ is the transition matrix of an arbitrary Markov
chain on the states $0,1,\dots,S$ and $\alpha_{ij}$ is given by
\begin{eqnarray}
  \alpha_{ij} = \frac{s_{ij}}{1+(\pi_i q_{ij})/(\pi_j q_{ji})}\,,
\end{eqnarray}
where it is noticeable that the normalization of the distribution
$\pi$ drops out. Metropolis updating $(s^M_{ij})$ is a special case:
\begin{eqnarray}
  s_{ij}^M = \cases{1+(\pi_i q_{ij})/(\pi_j q_{ji})\ {\rm for}\
  (\pi_jq_{ji})/(\pi_iq_{ij})\ge 1\ \Rightarrow\ \alpha_{ij}^M=1\,,
  \cr 1+(\pi_j q_{ji})/(\pi_i q_{ij})\ {\rm for}\ (\pi_jq_{ji})/
  (\pi_iq_{ij})< 1\ \Rightarrow\ \alpha_{ij}^M=(\pi_j q_{ji})/
  (\pi_i q_{ij})\,.}
\end{eqnarray}
It goes for $q_{ij}\ne q_{ji}$ sometimes under the name biased 
Metropolis updating. Requiring symmetry:
\begin{eqnarray}
  q_{ij} = q_{ji}~~\Rightarrow~~\alpha_{ij}^M =
  \cases{1~{\rm for}~\pi_j/\pi_i\ge 1\,,\cr
  \pi_j/\pi_i~{\rm for}~\pi_j/\pi_i< 1\,,}
\end{eqnarray}
and assuming the Boltzmann distribution $\pi_i=\exp(-\beta E_i)$, 
we obtain the original Metropolis algorithm back.

However, Hastings did not make progress in identifying suitable
distributions $\pi$ for important applications and error estimates 
of MCMC simulations are nowadays much better under control~\cite{BB04}. 
Besides, Hastings appeared to be unaware of Glauber's earlier work 
\cite{Glauber} generalizing the Metropolis approach.

\section{Umbrella sampling} \label{sec_umbrella}

Until about 1990 the paradigm of MCMC calculations in statistical
physics appeared to be that the use of Boltzmann weights will
optimize the simulation. However, there had been notable exceptions.

\begin{figure}
\begin{center}
\resizebox{0.95\columnwidth}{!}{%
  \includegraphics{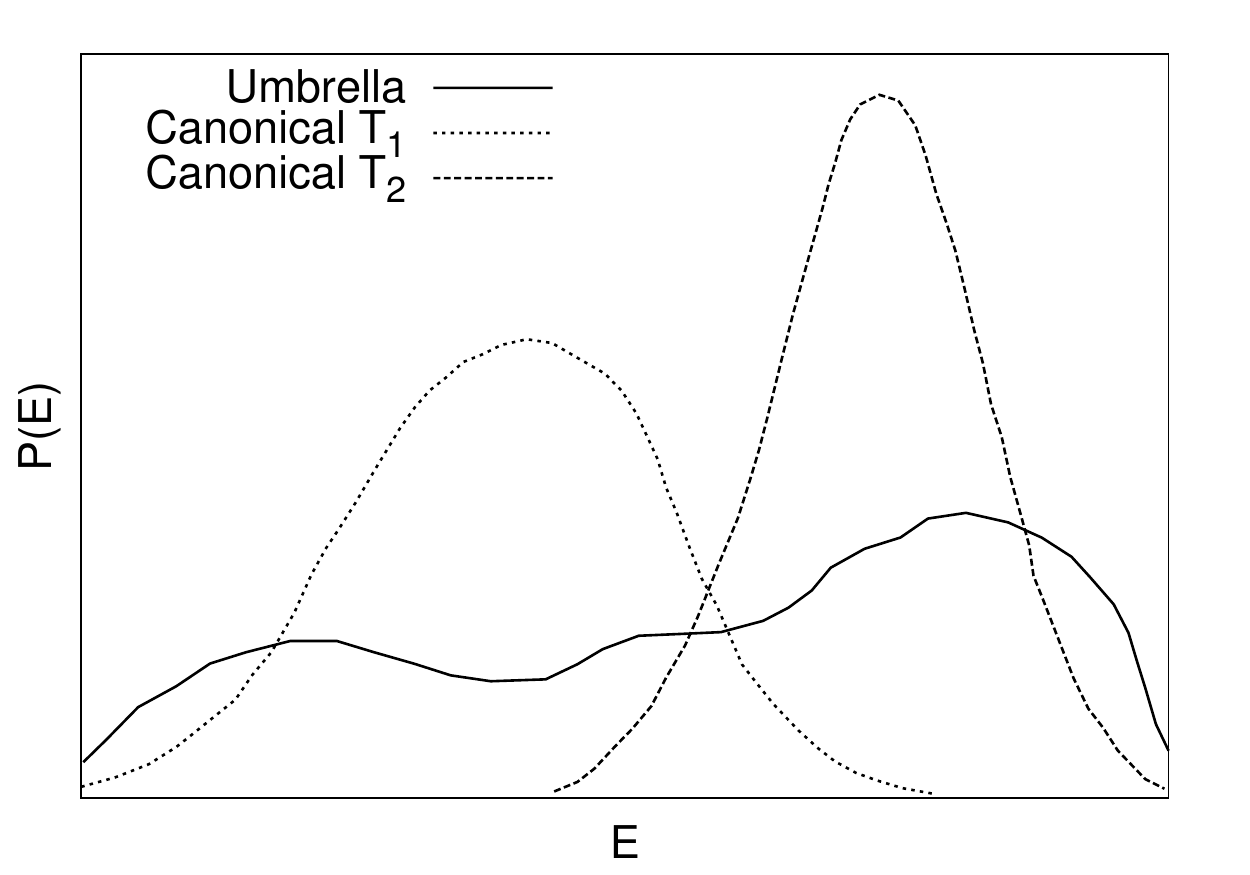} }
\end{center}
\caption{Probability density $P(E)$, $E$ energy, for a umbrella sample 
(solid line) together with canonical probabilities at two temperatures,
$T_1$ and $T_2$, as obtained by reweighting of the umbrella sample 
(sketched according to \cite{TV77}).} \label{fig_TV} \end{figure}

Already in 1972 Valleau and Card \cite{VC72} used overlapping bridging
distributions to estimate entropy and free energy by what they called
``multistage sampling''. Nowadays we call this multi-histogram 
reweighting. Torrie and Valleau \cite{TV74} continued with estimation 
of free energies from distributions designed for this purpose. Patey and
Valleau \cite{PV75} sampled ionic potentials with weights $w(r_{12})\,
\exp(-U/kT)$, where the weighting functions $w(r_{12})$ was chosen to 
spread the sampling over a desired range. In 1977 this line of work 
culminated in the proposal of ``Nonphysical Sampling Distributions in 
Monte Carlo Free-Energy Estimation: Umbrella Sampling" by Torrie and 
Valleau \cite{TV77}.

Markov chains, which sample these distributions, are special cases of 
Hastings framework, but the real problem is the engineering of sampling 
weights that work for a particular purpose. In any case, no reference 
to Hastings is given, though his \cite{Hastings} and the Valleau et al.\ 
papers \cite{VC72,TV74,PV75,TV77} have in common to be products of 
Toronto University.

The basic idea of umbrella sampling is sketched in Fig.~\ref{fig_TV}: 
The umbrella distribution is much broader than canonical distributions 
in its range, which are then not obtained by direct simulations, but by 
reweighting of a umbrella distribution. This leads us to the question 
``How to get the umbrella distribution in the first place?" and we find 
in the paper of Torrie and Valleau \cite{TV77}: 
``Superficially, the most serious limitation of the sampling techniques
described here may appear to be the lack of a direct and straightforward
way of determining the weighting function to use for a given problem. 
Instead $w({\bf q}^N)$ must be determined by a trial-and-error procedure
for each case, often beginning with the information available from the 
distribution in a very short Boltzmann-weighted experiment which is then
broadened in stages through subsequent short test runs with successively
greater bias of the sampling. What this rather inelegant procedure lacks
aesthetically is more than compensated by the efficiency of the ultimate
umbrella-sampling experiment." 

Umbrella sampling survived mainly in a niche of computational chemistry, 
e.g., \cite{Ch85}. But ten years later the conclusion by Lie and 
Scheraga \cite{LS88} was still: ``The difficulty of finding such 
weighting factors has prevented wide applications of umbrella sampling".

What is really gained by using umbrella distributions instead 
of a series of canonical samples and possibly patching them? The 
documentation at its time remained rudimentary and gains appear to
be no larger than, maybe, factors of 10 to 20. An exception is the 
improvement one can find in chapter~6.3 of the 1987 statistical 
mechanics book by Chandler \cite{Chandler}. He considered a $20\times 
20$ Ising model below the (critical) Curie temperature, where 
visitation of states with
magnetization $m=0$ is a rare event. By patching 10 umbrella windows 
he can reach $m=0$ and sample configurations which are in canonical 
Boltzmann simulations suppressed by 6 to 7 orders of magnitude. 
Unfortunately, this impressive demonstration of the numerical 
method missed a physical focus. Naturally this would have been a
calculation of the interface tension, known exactly since Onsager's
\cite{Onsager} 1944 solution of the 2D Ising model. However, this
requires Binder's \cite{Bi82} relation (\ref{fsL}), which is discussed 
in the next section (presumably Chandler was not aware of it).

Independently of Chandler, and unaware of umbrella sampling, Bhanot 
and collaborators \cite{Bh87} developed a similar method of patching 
microcanonical simulations, then defined as simulations restricted to 
some small energy range $E_{\min}\le E\le E_{\max}$, and applied it 
to a number of physical topics.

\section{Multicanonical and other generalized ensembles} \label{sec_muca}

The spread of the multicanonical approach is an example of an 
interdisciplinary success story with unexpected turns of events. 
In 1988 Parisi et al.\ \cite{Pa88} claimed that the deconfining phase 
transition of SU(3) LGT is second order, what appeared rather weird, 
because a large number of previous publications had agreed on a weak 
first order transition. However, as this was Georgio Parisi, everyone 
felt obliged to take this very seriously and immediately, as well as 
over the subsequent years, a large number of papers appeared, including 
one by myself and collaborators \cite{BB90}, which all re-confirmed the 
first order nature of this transition. 

This also stimulated work on simpler models than SU(3) LGT with the
aim to develop better methods to identify first order transitions. 
Examples of simple systems with first order phase transitions are 2D 
and 3D Ising models below the Curie temperature for which the transition 
is at $h=0$ in an applied magnetic field and $q$-state Potts models for 
which there is a first order transition in the temperature ($q\ge 4$
in 2D, $q\ge 3$ in 3D required). One focus was on calculations of their 
interface tensions. In particular, Potvin and Rebbi \cite{PR89} as well 
as Kajanti et al.\ \cite{Ka89}, used the 7-state 2D Potts model as 
laboratory for a newly developed approach, which estimated the interface 
tension by driving half a lattice ``adiabatically'' from one phase into 
the other. Thomas Neuhaus and I stumbled in this context over Binder's 
method for estimating interface tensions.

\subsection{Binder's method}

\begin{figure} \begin{center}
\resizebox{1.00\columnwidth}{!}{%
  \includegraphics{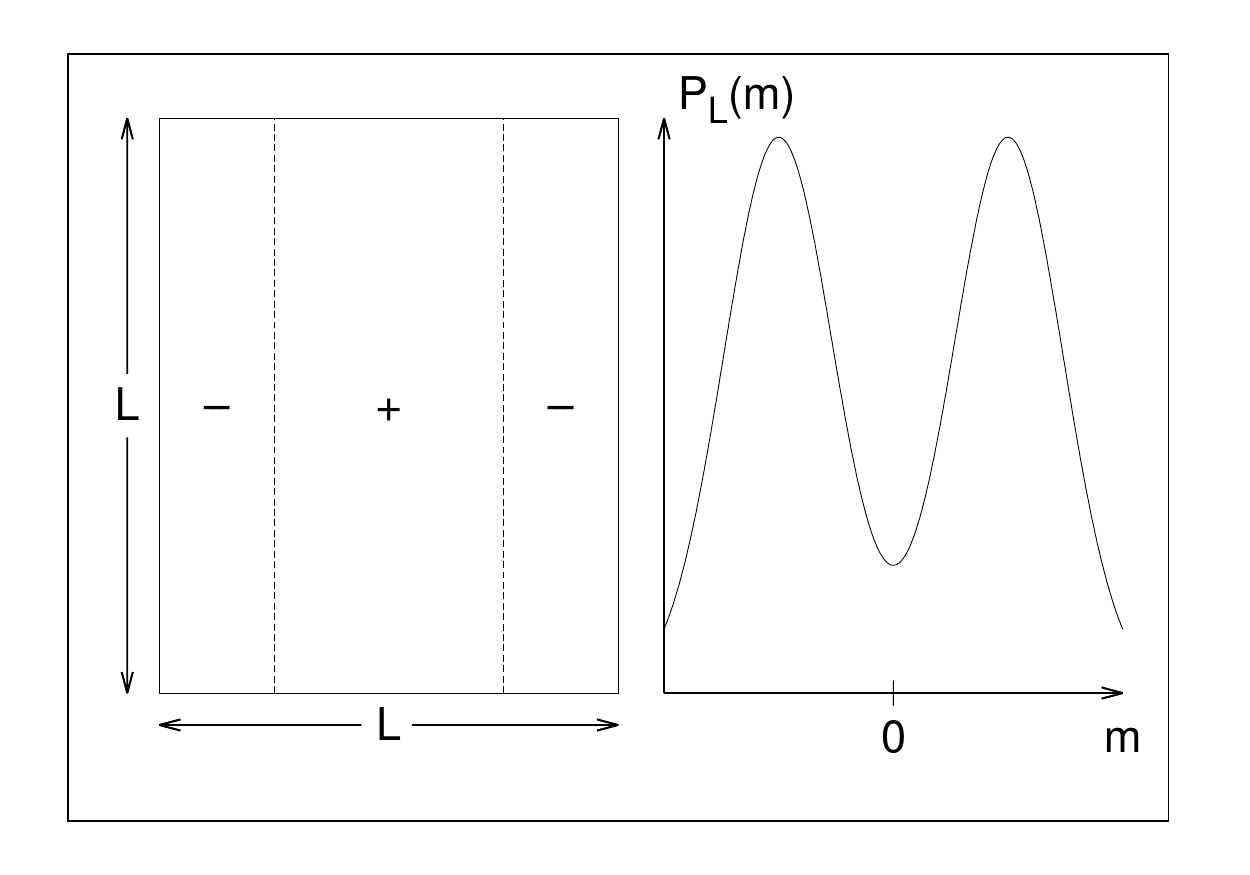} }
\end{center}
\caption{Ising model with periodic boundary conditions below the Curie 
temperature: Order parameter (magnetization) distribution (qualitatively
drawn after Ref.~\cite{Bi82}).} \label{fig_Binder} \end{figure}

Figure~\ref{fig_Binder} sketches the probability density of the 
magnetization $m$ for a 2D Ising model with periodic boundary conditions 
at a temperature below its critical and at external magnetic field 
$h=0$. For the interface tension $f^s$ Binder \cite{Bi82} derived the 
relation 
\begin{eqnarray} \label{fsL}
  f^s\ =\ \lim_{L\to\infty} f^s_L~~{\rm with}~~ f^s_L\ =\ - \frac{1}
  {L^{(D-1)}}\,\ln\left(\frac{P^{\min}_L}{P^{\max}_L}\right)\,,
\end{eqnarray}
where $P^{\max}_L$ is the maximum and $P^{\min}_L$ the minimum of the
probability density of the magnetization on a finite lattice of size
$L^D$ with periodic boundary conditions. In case of the order-order
transition of Fig.~\ref{fig_Binder} at $h=0$ it is obvious that the 
probability density is symmetric in the magnetization $m$. For a 
temperature driven first order transition there is no such symmetry. 
The energy, instead of the magnetization, is recorded on the abscissa 
and it becomes part of the calculations to determine suitable 
pseudotransition temperatures $\beta_t(L)$.  

On finite lattices $P^{\max}_L$ and $P^{\min}_L$ can be calculated by 
MCMC, but initial attempts \cite{Bi82} remained pitiful, because they 
were carried out in the canonical ensemble and estimates 
deteriorate then quickly with increasing lattice size as configurations 
for $P^{\min}_L$ are exponentially suppressed according to $P^{\min}_L
\sim\exp\left(-f^sL^{D-1}\right)$. So, Binder's method remained dormant
for almost ten years.

\subsection{Multicanonical ensemble}

\begin{figure}
\begin{center} \resizebox{1.00\columnwidth}{!}{%
  \includegraphics{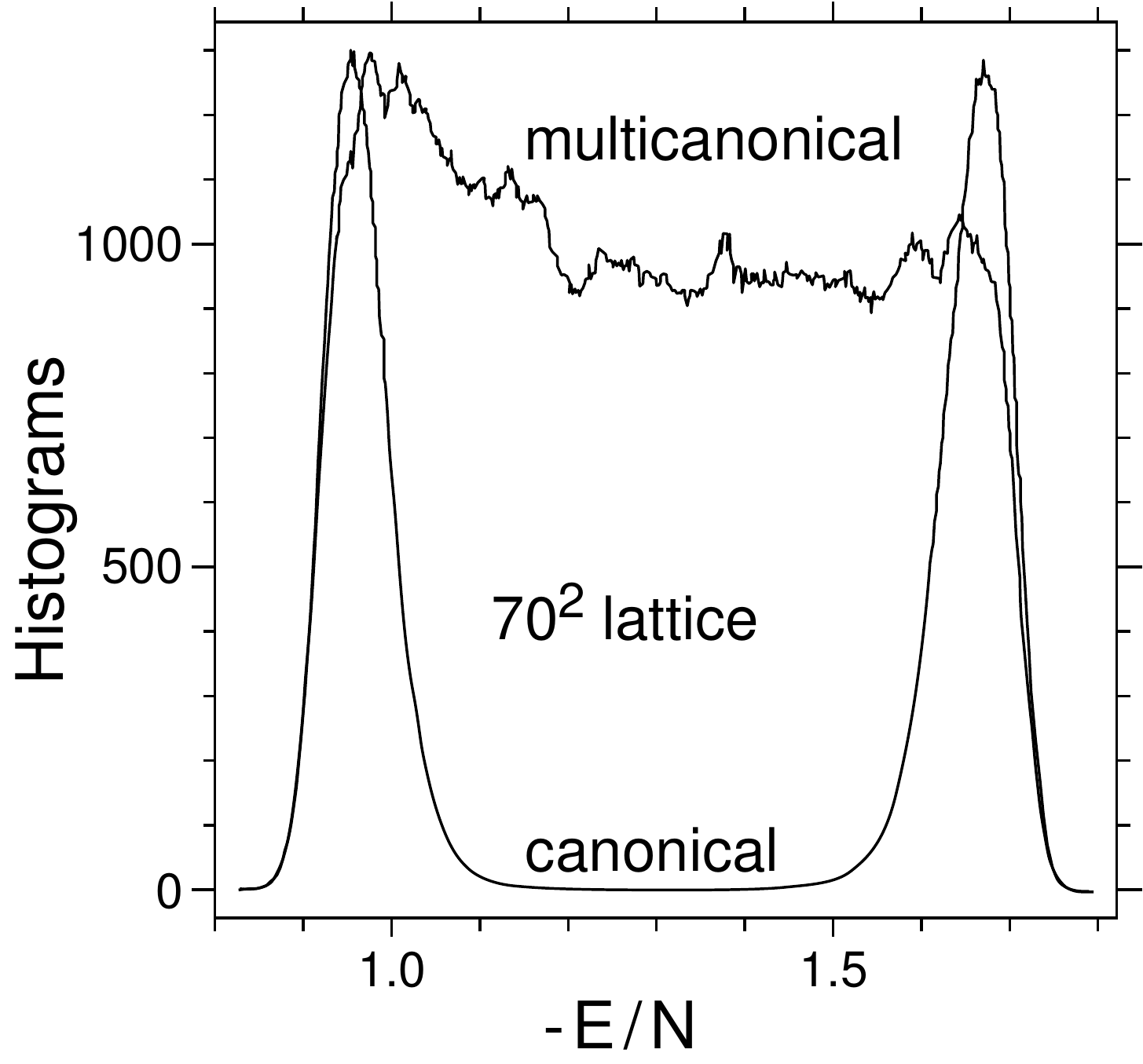} }
\end{center}
\caption{Reweighting of a MUCA simulation at the first order phase
transition of the 2D 10-state Potts model to the canonical ensemble
\cite{BN92}.} \label{fig_BN01} \end{figure}

\begin{figure} \vskip -90pt \hskip -33pt
\resizebox{1.18\columnwidth}{!}{\includegraphics{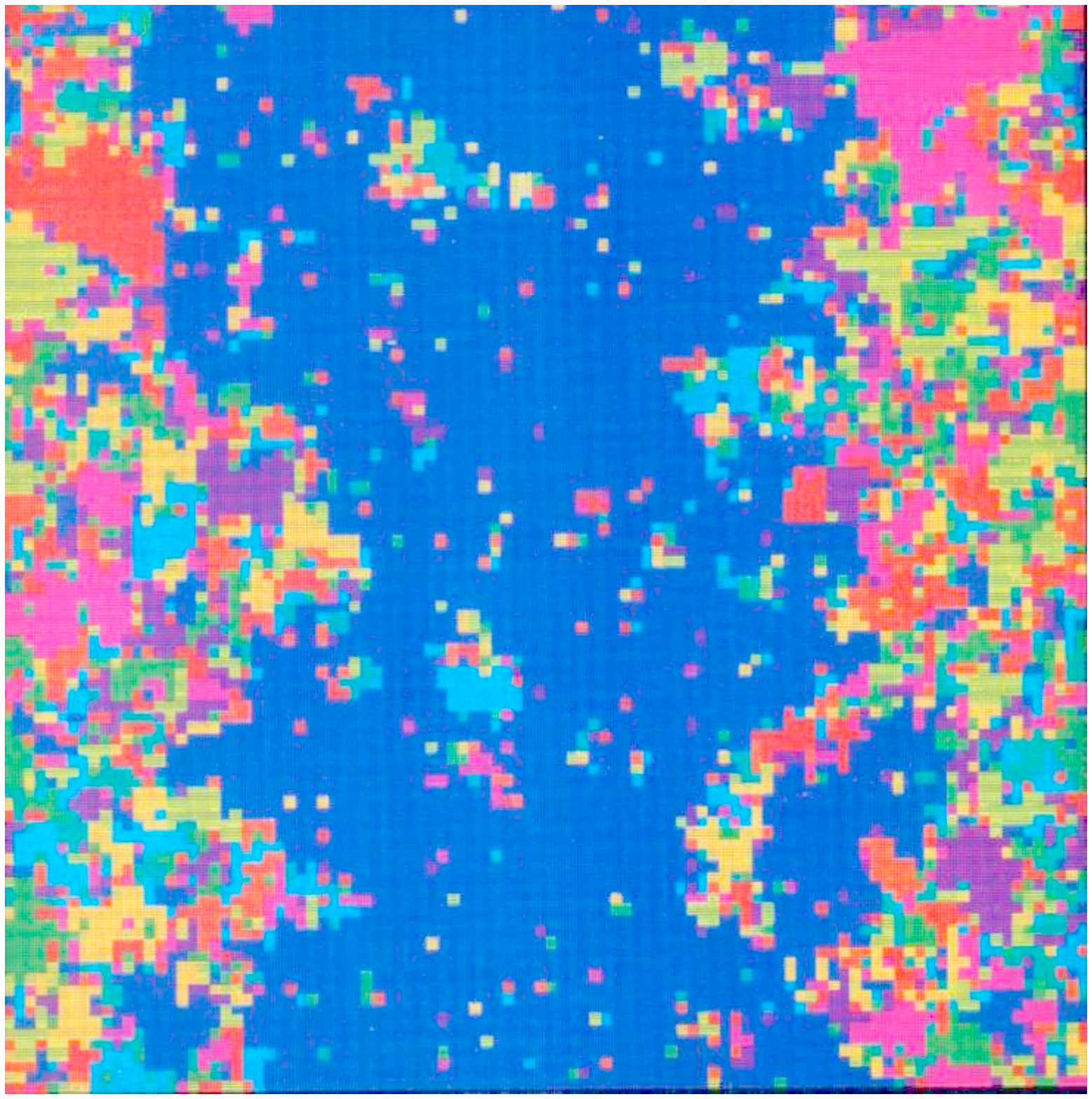}}
\vskip -90pt
\caption{2D 10-state Potts model: A configuration with disorder-order
interfaces from the $P_L^{\min}$ configurations of the canonical 
ensemble.} \label{fig_colorBN} \end{figure} 

For a temperature driven first order transition the MUCA approach 
\cite{BN91,BN92} calculates $P_L^{\min}$ by sampling no longer with 
Boltzmann weights but with the inverse number of states, 
\begin{eqnarray} \label{WMUCA}
  W(E)\ =\ {\rm const}/n(E)\, ,
\end{eqnarray}
as weights. Subsequently, one reweights the thus obtained MUCA ensemble 
to the canonical ensemble at an effective transition temperature, 
$\beta_t(L)$, for which equal heights of the maxima of the spectral 
probability density are realized (equal weights have also been 
suggested \cite{BK92}).

In contrast to umbrella sampling the MUCA weights are precisely defined,
but in principle MUCA appears to suffer from the same shortcoming: The
weights are a-priory unknown (in contrast to the Boltzmann weights).
However, well defined target weights led to a clear focus on the 
problem of getting them. Especially, for the envisioned calculations 
of interface tensions weights can be obtained by finite size 
extrapolations from small to increasingly larger systems: For very 
small systems canonical simulations work and provide the starting point. 
Then one extrapolates weights for the next larger system, corrects them 
by simulation results from this system, proceeds to the next larger 
system, and so on.

\begin{figure}
\begin{center} \resizebox{1.00\columnwidth}{!}{%
  \includegraphics{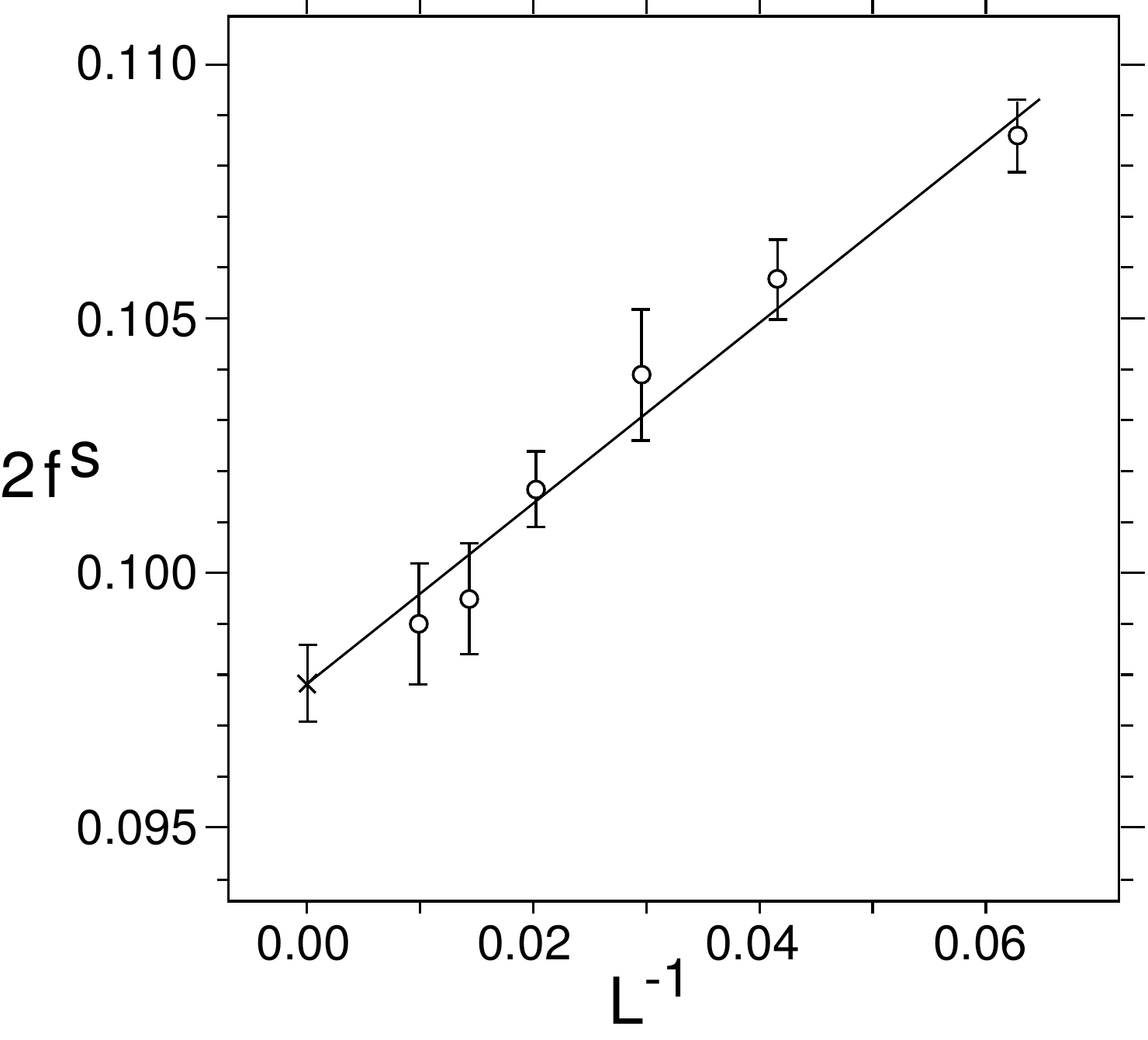} }
\end{center}
\caption{Extrapolation of the interface tension for the 2D 10-state
Potts model~\cite{BN92}.} \label{fig_BN04} \end{figure}

\begin{table}
\caption{Interface tensions $f^s$ for some 2D Potts models.}
\label{tab_fs} \begin{tabular}{llll}
\hline\noalign{\smallskip}
$q$ & MUCA \cite{JBK92,BN92} & Potvin and Rebbi \cite{PR89} 
& Exact (Borgs and Janke \cite{BJ92})  \\
\noalign{\smallskip}\hline\noalign{\smallskip}
 7 & 0.0241\ (10) & 0.1886 (12) & 0.020792$\dots$ \\
10 & 0.09781 (75) & $-$         & 0.094701$\dots$ \\
\noalign{\smallskip}\hline
\end{tabular} \end{table}

For the 2D 10-state Potts model lattices up to size $100^2$ were covered
in this way. Figure~\ref{fig_BN01} shows for a $70^2$ lattice the energy 
histogram obtained by using approximate MUCA weights and at the 
pseudotransition temperature the canonically reweighted histogram. 
Improvements by a factor $10^5$ over canonical simulations are easily 
achieved. Figure~\ref{fig_colorBN} gives an example of a configuration
with interfaces as found for $P^{min}_L$ in the canonical ensemble. 
Equation (\ref{fsL}) yields for each lattice size the effective
interface tension and in Fig.~\ref{fig_BN04} these estimates are 
extrapolated to the infinite volume result ($1/L\to 0$). The thus 
obtained $f^s$ is the second entry of the third row of Table~\ref{tab_fs}.

However, there was a problem now: The MUCA estimate for the interface 
tension of the 2D 10-state Potts model is lower than this estimate for 
the 2D 7-state Potts model by Potvin and Rebbi \cite{PR89}, the third 
entry of the second row of Table~\ref{tab_fs} (Kajanti et al.\ 
\cite{Ka89} agree and give $2f^s\approx 0.2$ without error bar). As 
the strength of the 2D Potts model transitions increases with the number 
of states, this should be the other way round. The discrepancy was 
confirmed by a direct MUCA calculation of the 2D 7-state Potts model
\cite{JBK92}\footnote{This paper was actually my first collaboration
with W.\ Janke, who was then on a joint postdoc appointment with the 
Embry Riddle Aeronautical University at Daytona Beach and the Florida 
State University.} for which $f^s$ is listed as the second entry of the 
second row of Table~\ref{tab_fs}. What now? As nobody will easily admit 
that his or her numerical method does not work, an endless debate was 
looming. But:

\subsection{Miracles happen (thanks to Borgs and Janke)}

After the numerical results were published Borgs and Janke
\cite{BJ92} realized, based on previous work by Buffenoir and Wallon 
\cite{BW92} and others, that there is actually an exact solution 
for the interface tension of 2D $q$-state Potts models. While numerical 
calculations tend to reproduce known exact results accurately, this was 
not so much the case for these previously unknown exact results, which 
are for the 2D 7-state and 10-state models collected in the 4th column 
of Table~\ref{tab_fs}. The MUCA estimate for the 10-state Potts model 
agrees within less than 4\%, but only within five sigma. For the 7-state
model the MUCA estimate needs three error bars to include the exact 
value. However, this is very good when compared to the other numerical 
estimate for the 7-state model, which is an entire order of magnitude 
too large. So, due to the paper by Borgs and Janke the looming endless 
discussion was cut short (the interfaces are perceived as too stiff in 
the method of \cite{PR89,Ka89}). Looking back, the controversy helped 
a lot to draw attention on MUCA simulations as a reliable method. Almost 
needless to say, once the exact results are known, the discrepancies 
with MUCA estimates can be resolved by applying more sophisticated fits 
\cite{BNB94} than the one of Fig.~\ref{fig_BN04}.

\subsection{Other early multicanonical simulations and new horizons}

\begin{figure}
\vskip -80pt
\begin{center} \resizebox{1.00\columnwidth}{!}{%
  \includegraphics{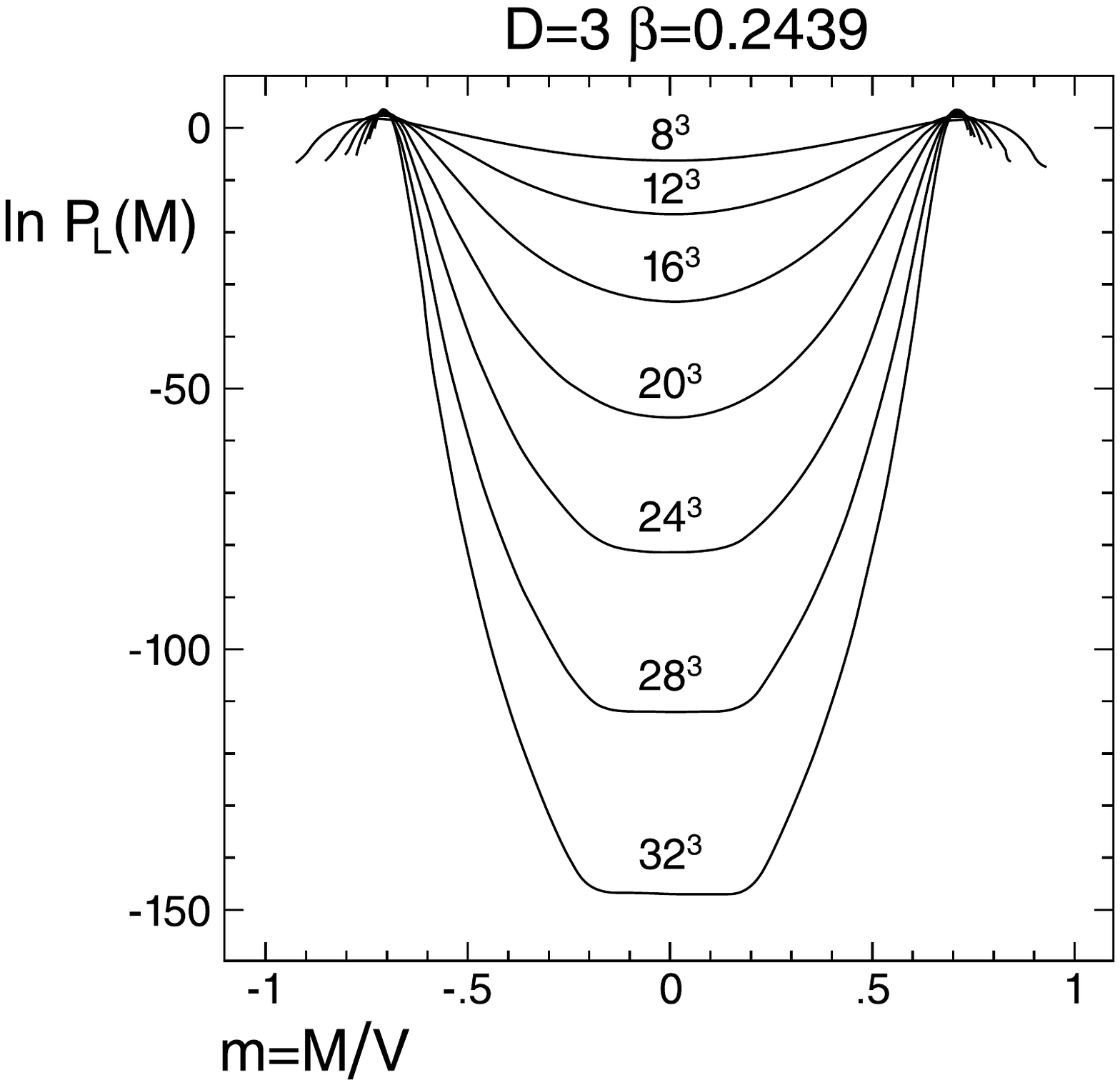} }
\end{center}
\vskip -99pt
\caption{Probability densities of the magnetization for the 
3D Ising model at a temperature below its critical value
\cite{BHN93}.} \label{fig_BHN} 
\begin{center} \resizebox{1.00\columnwidth}{!}{%
  \includegraphics{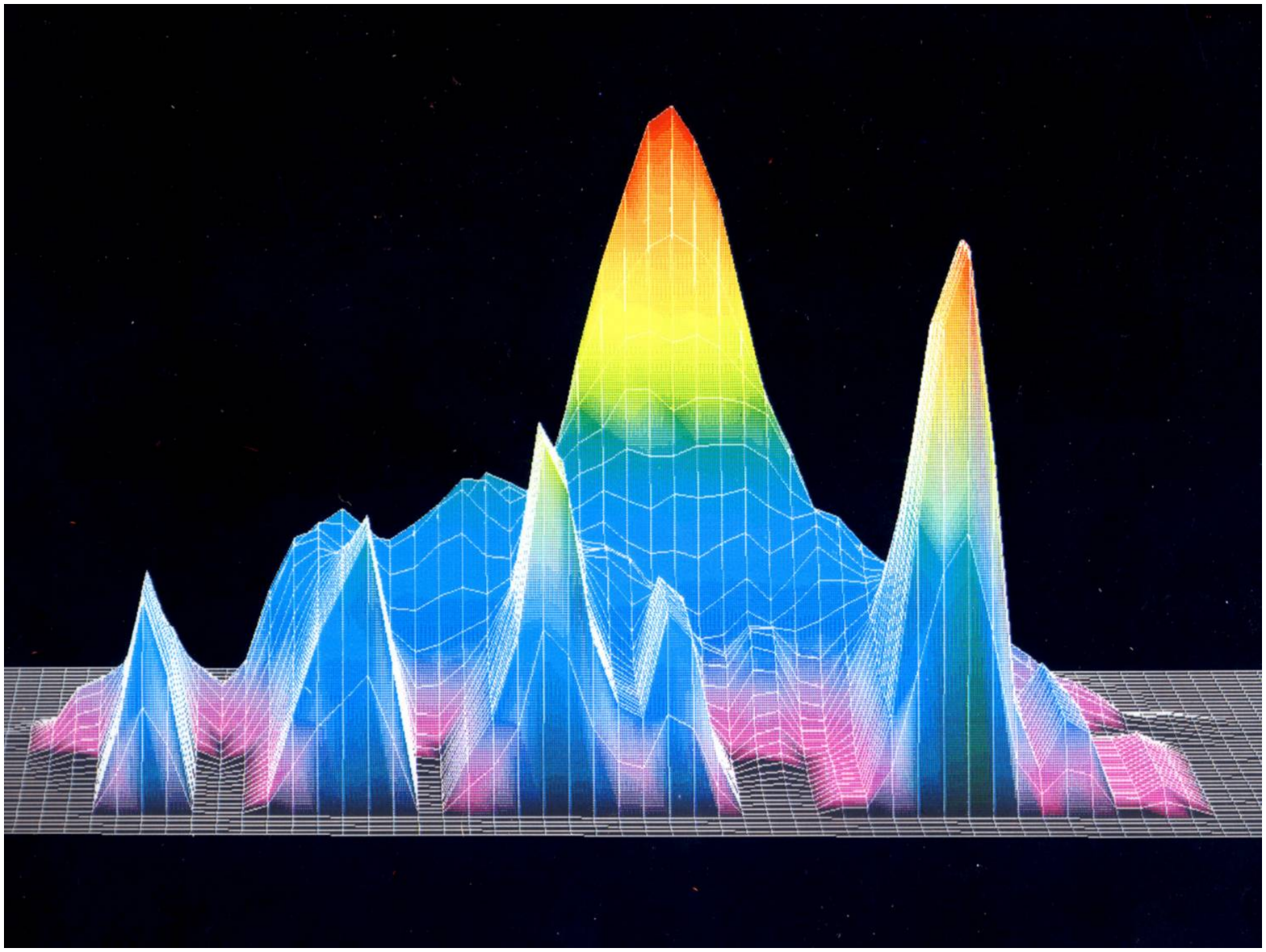} }
\end{center}
\caption{Snapshot from a MUCA simulation of the 2D Edwards-Anderson 
Ising spin glass, which exhibits barriers in the overlap parameter 
\cite{BC92}.} \label{fig_spg} 
\end{figure}

Soon MUCA calculations were also carried out for the interface tensions
of the order-order transitions of 2D and 3D Ising models below their 
critical temperatures and MUCA is in this application called 
multimagnetical (MUMA). The 2D case allows for comparison with Onsager's 
\cite{Onsager} exact result and MUMA calculations were well consistent
\cite{BHN91}. This was certainly of interest when this paper was 
submitted in September 1991 and the controversy about the estimates for 
Potts models was still unsettled. However, due to the malice of some 
referees, which were finally overruled by the editor, the paper appeared  
only in January 1993. By then subsequent work \cite{BHN93}, which 
included estimates for 3D Ising model interface tensions, appeared 
almost simultaneously in print. Figure~\ref{fig_BHN} shows the 
reweighted canonical probability densities of one of our MUMA 
simulations of the 3D Ising model and it is clear that an astronomically 
large improvement over canonical simulation is reached: Configurations 
for the minimum are sampled, which are suppressed by a factor of 
approximately $e^{-150}\approx 10^{-60}$ in the canonical ensemble. 

Somewhat later extensions of the MUCA method to cluster algorithms, 
multibondic (MUBO) simulations, were introduced and studied by Janke 
and collaborators~\cite{JK95,BJ07}.

Realizing the importance of the MUCA approach beyond first order phase 
transitions, Celik and I \cite{BC92} published in 1992 an 
application to complex systems with a rugged free energy landscape 
due to frustrated interactions: The 2D Edwards-Anderson Ising spin 
glass. By sampling with MUCA weights (\ref{WMUCA}) one can move in 
and out of valleys and find their relative heights by connecting them 
through the disordered phase. This is illustrated in Fig.~\ref{fig_spg}: 
The mountain in the background is sampled from configurations of the 
disordered high temperature phase and the MUCA updating process travels 
in and out of low temperature valleys (this is where the histogram 
entries are), which are separated by high barriers (means no histogram 
entries) in the probability density of the overlap parameter. Our spin 
glass investigations with MUCA went on for quite a while \cite{BCH} and 
the final paper with Billoire and Janke \cite{BBJ} was about overlap 
barriers in the 3D and 4D Edwards-Anderson Ising spin glass model. In 
contrast to first order phase transitions finite size extrapolations of 
the MUCA weights are not possible for spin glasses, because each replica 
is unique. So, Ref.~\cite{BC92} introduced a recursive approach to 
estimate them. Progress has since then been achieved with respect to 
this and is shortly reviewed in section~\ref{subsec_WL}.

\begin{figure}  \vskip -160pt 
\begin{center} 
\resizebox{1.10\columnwidth}{!}{\includegraphics{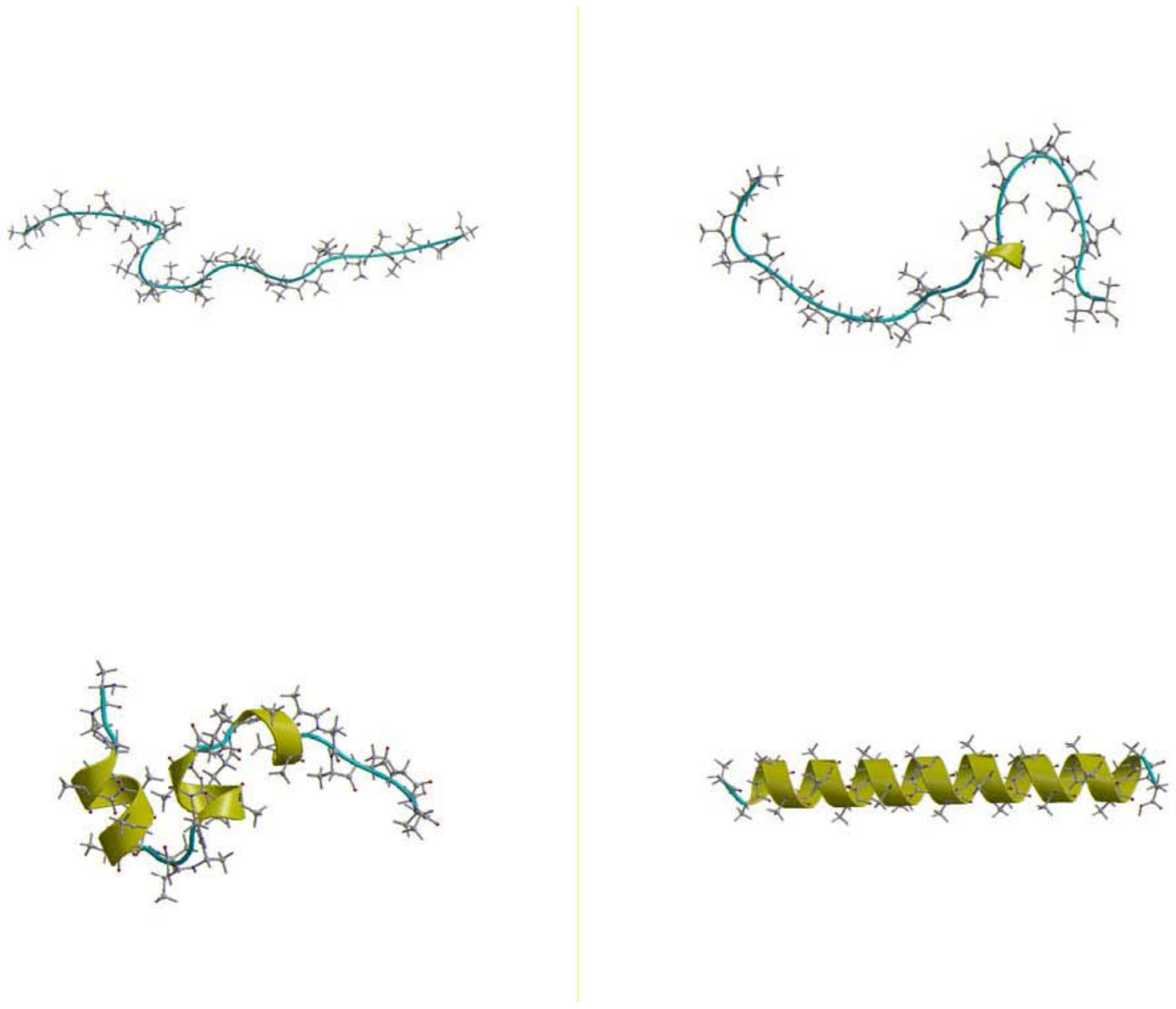}}
\end{center}  \vskip -160pt
\caption{Configurations from folding of poly-alanine into its helix 
groundstate \cite{HO99a}.} \label{fig_alahel} \end{figure}

The concluding remarks in \cite{BC92} are: ``The similarities of spin 
glasses to other problems with conflicting constraints suggest that 
MUCA simulations may be of value for a wide range of investigations:
optimization problems like the traveling salesman, neural networks, 
protein folding, and others.'' Hansmann and Okamoto made the application 
to small proteins reality \cite{HO93}. Figure~\ref{fig_alahel} pictures 
the folding of poly-alanine configurations into their helix groundstates 
in a MUCA simulation that connects the native groundstate with a 
disordered initial configuration \cite{HO99a}. Their review article 
\cite{HO99b} coined the generic name ``Generalized Ensembles'' for 
MUCA ensembles, umbrella sampling, expanded ensembles and the replica 
exchange method.

\subsection{Expanded ensembles and the replica exchange method}

Expanded ensembles \cite{LM92,MP92} enlarge the configuration space 
by one or more additional variables. If this is the temperature, the 
method is also called ``simulated tempering''. In the replica exchange 
method \cite{Ge91,HN96} several identical replica of the systems are 
simulated with parameter values chosen so that transitions between the 
replica, which are accepted according to the Metropolis method, become 
possible, i.e., have reasonable acceptance rates. This is a special 
case of the approach of Ref.~\cite{SW86}, which was not focused enough 
to provide guidance for practical applications. The replica exchange 
method allows for easy parallelization, because the simulation of each 
replica is independent from that of the other replica and the exchanges 
requires only transfer of a few variables. See \cite{Ja08} for work by
Janke and collaborators on this subject.

If the replica differ in the temperature variable, the replica exchange 
method is often called ``parallel tempering'' (PT), which should not be 
confused with simulated tempering. Presumably due to the easy use of 
many processors on supercomputers, PT has become quite 
popular, in particular for protein simulations to which it was introduced 
in Ref.~\cite{Ha97} within the MCMC framework. However, in biophysics 
molecular dynamics (MD) simulations are most widely used. To them 
PT was adapted a few years later in Ref.~\cite{SO99}.

That all these methods flourished since the early 1990s, and were not
forgotten again, shows that by now the time was right to create a 
sufficiently large and sophisticated user community.

\subsection{Performance of generalized ensemble simulations}

\begin{figure} \begin{center} 
\resizebox{1.00\columnwidth}{!}{
\includegraphics{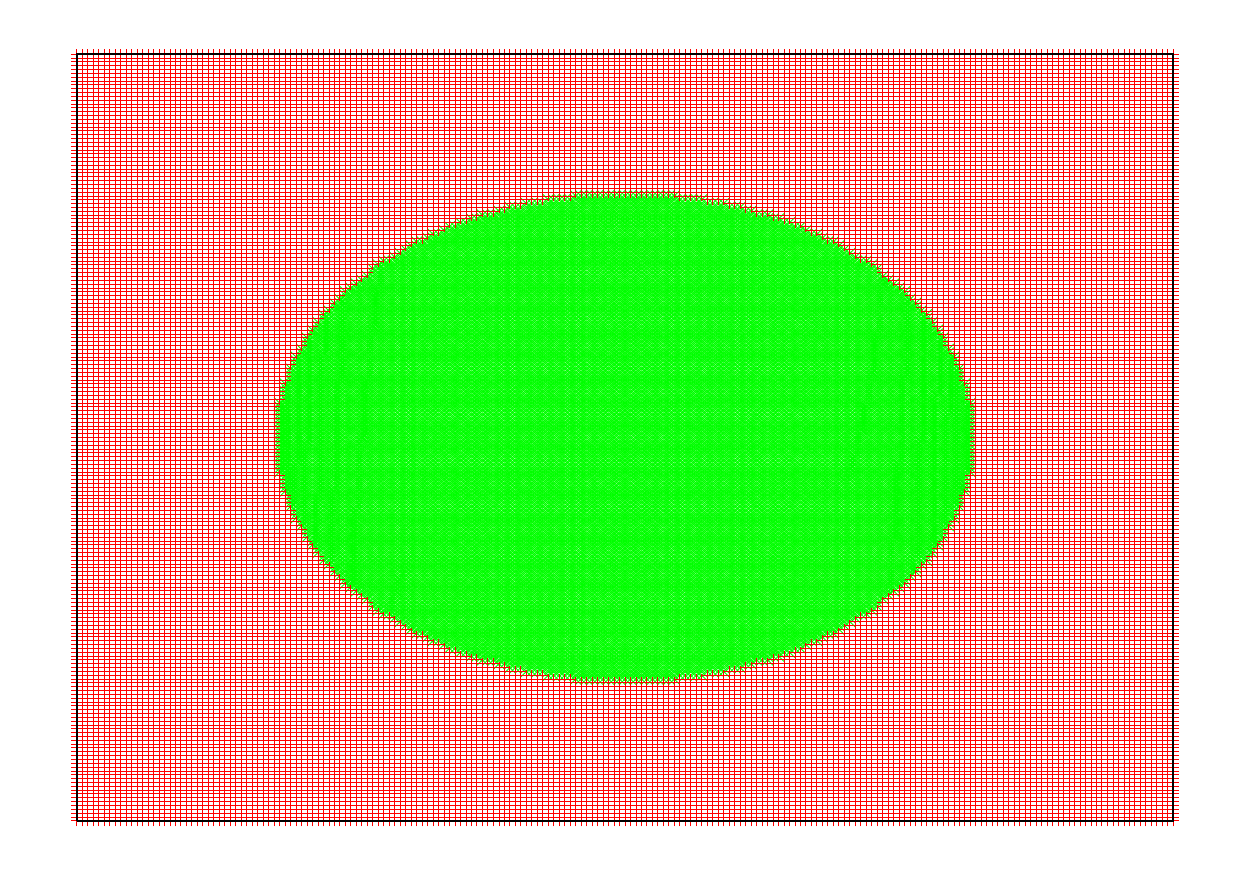}\hfill
\includegraphics{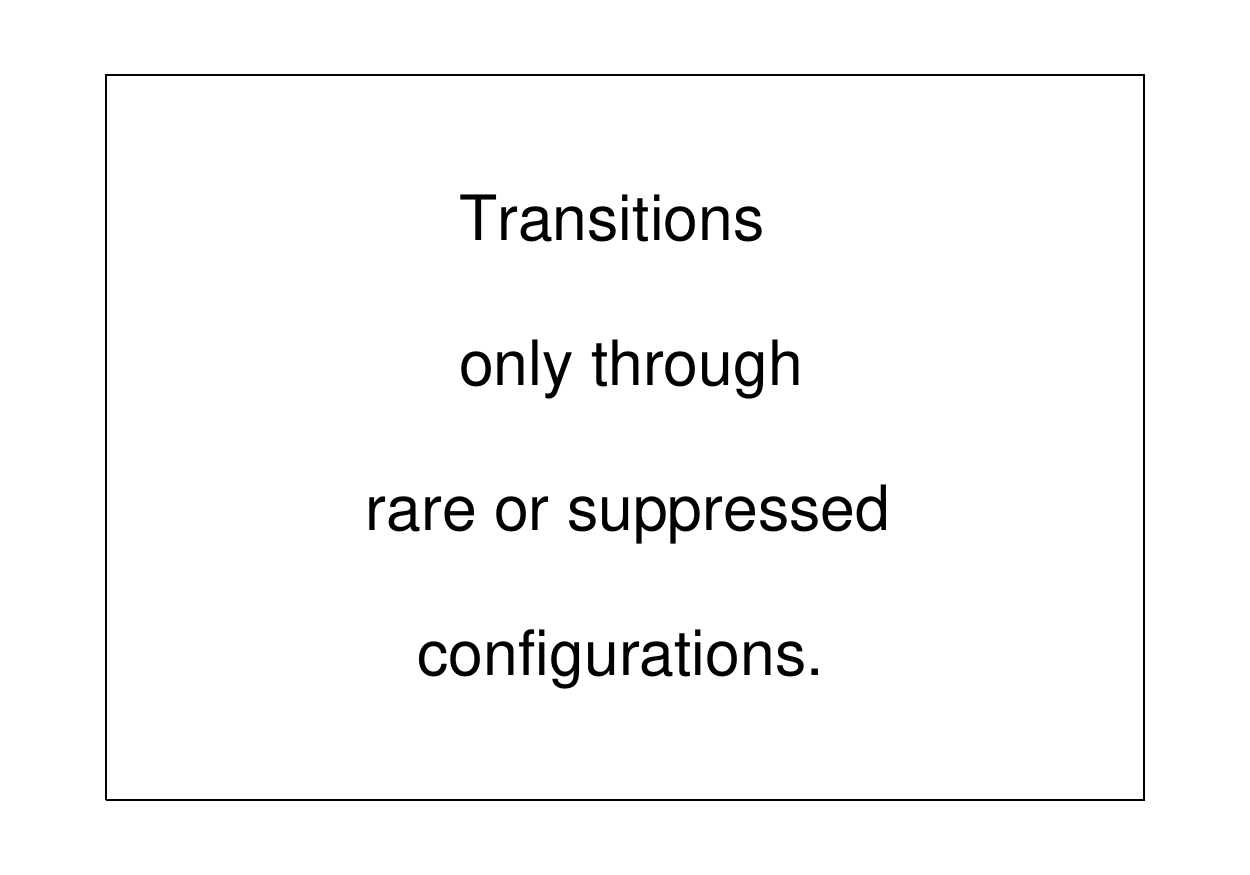}\hskip -0.07in
\includegraphics{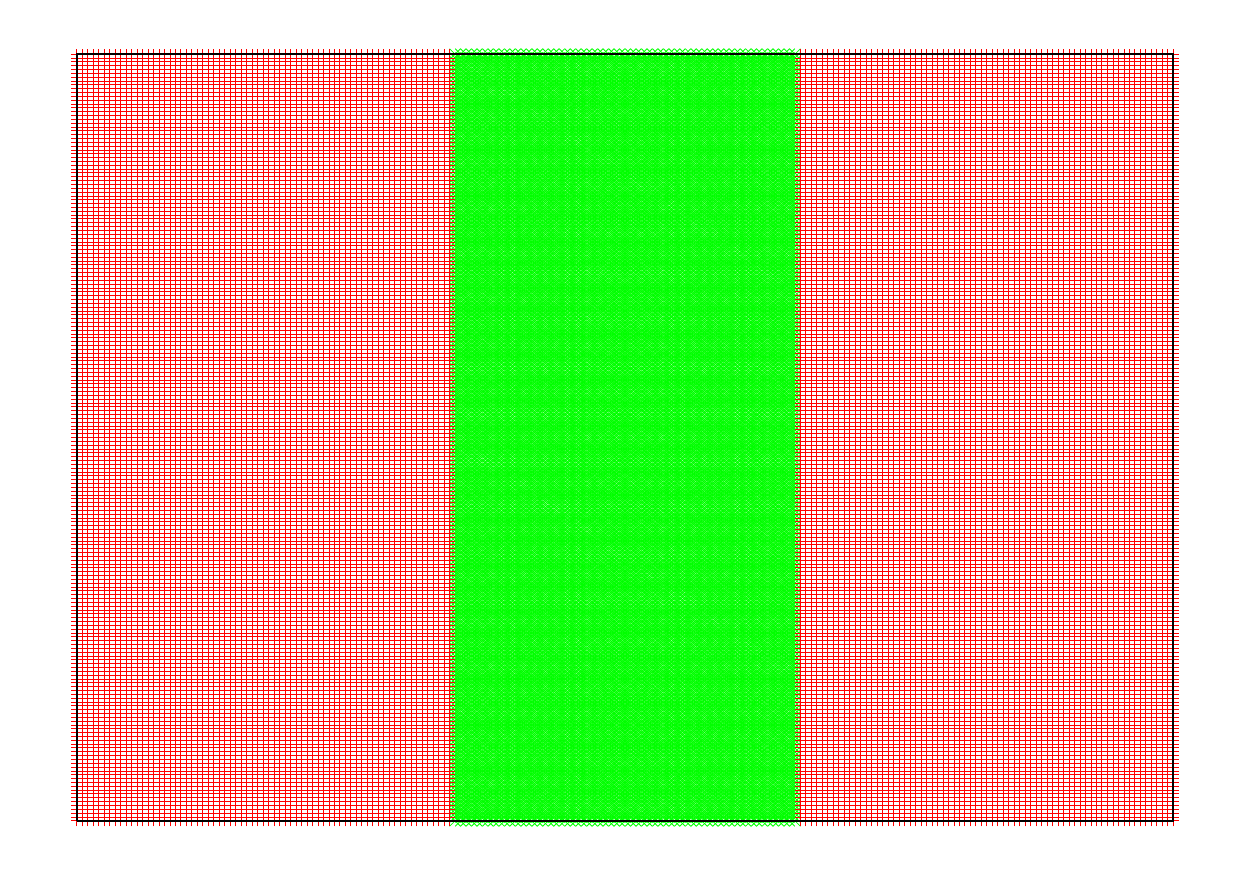} }
\end{center}  
\caption{Energy optimized droplet versus percolated configuration
as discussed in \cite{NH03}.} \label{fig_NH} \end{figure}

The number of steps a random walk (diffusive process) needs to bridge 
a distance $N$ is proportional to $N^2$. Therefore, the optimal 
performance of a MUCA simulation to bridge an extensive energy gap 
is proportional to $V^2$, where $V$ is the volume of the system. 
In the early work \cite{BN92} it was assumed that for first order 
transitions the performance would be close to this optimum, something 
like $V^{2.3}$ was consistent with the data. This turned out to be too 
optimistic. The slowing down is still exponential in system size $L$. 
Only the worst exponential, $\sim\exp\left(-f^sL^{D-1}\right)$, has 
been eliminated. The argument due to Neuhaus and Hager \cite{NH03} 
is illustrated in Fig.~\ref{fig_NH} for an order-order transition. The 
interface energy minimized shape for the green within the red phase is 
a droplet as long the green phase region is sufficiently small. 
However, once the green phase is sufficiently large, the optimal shape 
becomes the percolated green rectangle and transitions from the droplet
to the rectangle have to go through exponentially suppressed 
configurations. This is an example of hidden barriers for which 
a good reweighting variable is difficult to identify. In practice
the coefficient of this subleading exponential appears to be rather
small, so that the diffusive process dominates the slowing down for
some while and effectively a power law in $L$ is observed as long
as the systems are not too large.

This argument does not apply directly to PT, because
each replica is then simulated in the canonical ensemble. As a 
consequence rare configurations of free energy barriers like the
one of the right side of Fig.~\ref{fig_NH} are in practice not 
sampled but at best jumped and it is wrong to use this method
when one is interested in sampling barrier configurations. On the
other hand, when one is interested in connecting distinct groundstate
branches, like those of our spin glass Fig.~\ref{fig_spg}, through
the disordered phase, PT appears to be well suited.

\subsection{Wang-Landau recursion and parallelization} \label{subsec_WL}

Ever since \cite{BC92} recursions for MUCA weights received 
appropriate attention. The proposed schemes, like \cite{Be96},
relied on manipulations of additive histograms of the already 
covered energy ranges. This changed when Wang and Landau \cite{WL01} 
introduced a new sampling method that relies on a multiplicative 
instead of a additive recursion of weights:
\begin{eqnarray} \label{WL}
  \ln W(E)\ \to\ \ln W(E) - a\,,~~a>0\,,
\end{eqnarray}
whenever the (discrete) energy $E$ is sampled. Then $a\to a/2$ after
a sufficiently flat histogram is obtained. Though proposed as a
sampling method, in the MUCA context it is best used as a recursion
where the weights get frozen once they are working estimates of the 
MUCA weights. Here ``working estimate'' means that the energy range 
of interest is cycled with the frozen weights (instead of getting 
stuck in part of that range).
In tests, which I performed iterating the MUCA weights this way,
it has turned out to be a robust way of getting working estimates
within a small fraction of the CPU time needed for the subsequent 
calculation with fixed weights. So, there is little incentive to hunt 
for further improvement, although it is not finally settled whether 
additive recursions can do as well or even better (a poster \cite{Ja16}
comparing Wang-Landau and MUCA recursions was presented at this 
conference).

With Bazavov \cite{BB09} I have worked out a straightforward 
implementation of the Wang-Landau recursion for the case of a continuous 
energy variable. The basic steps are:
\begin{enumerate}
\item Bin into histograms just large enough so that a single updating 
step cannot jump a bin.
\item Track the mean values within each bin.
\item Do not use constant weights over a bin, but interpolate 
logarithmic weights (\ref{WL}) weights linearly between mean values of
neighbor bins.
\end{enumerate}
The performance was as good as the one achieved in models with discrete 
energies.

For simulations on supercomputers parallelization is crucial. Here 
I like to point the reader towards two relatively recent papers.
Massively parallel Wang-Landau sampling is discussed by Landau and
collaborations in \cite{La13} and parallelization of MUCA simulations 
by Janke and collaborators in \cite{Ja13}.

\section{Summary and conclusions} \label{sec_sum}

A message is that  mastering simulational methods is not the trivial 
part of a physical, chemical or  biophysical MCMC study. Astronomically 
large efficiency factors can float around between getting a simulation
optimized or not. There have been hundreds of papers refining and 
improving the methods sketched here. Within this limited review I 
cannot follow up on them. An efficient way to track such investigations 
is to look for citations of the papers quoted here and work from there 
up to the latest developments.

Obviously, for a number of applications generalized ensembles allowed 
great leaps in the efficiency of MCMC simulations, but ultimately it
is unavoidable that one reaches some limits where one is stuck again.
The two main problems appear to me:
\begin{enumerate}
\item There are often hidden barriers and we have been unable to 
find reaction coordinates in which these barriers become explicit. 
Recommendation: Keep trying.
\item The optimal performance of the discussed methods is limited 
by the $V^2$ power law due to being a diffusive process. For large
systems that can still be far too slow. 
\end{enumerate}
At second order phase transitions collective updating with cluster 
algorithms provides great improvements, but they have a rather limited
applications range. Non-equilibrium simulations could be a way out if 
one could relate the generated configurations to those of canonical 
equilibrium ensembles. Putting in some dynamics into simulations by
combinations of MCMC with MD may help. Most important, we are awaiting 
the not yet existing ideas of the next generation of computational 
physicists. 
\medskip

{\bf Acknowledgments:} I would like to thank Martin Weigel and the 
other organizers for inviting me to Wolfhard's belated birthday party.

\end{document}